\documentclass{cimento}

\usepackage{graphicx}  

\title{First measurements on charmless $B$ decays at Belle II}
\author{S.~Raiz\thanks{On behalf of the Belle II collaboration.}}
\instlist{
    \inst Dipartimento di Fisica dell'Università di Trieste e INFN Sezione di Trieste — Trieste, Italy}

\begin{document}

\maketitle

\begin{abstract}
We report on the first Belle II measurements of branching fractions and CP-violating charge asymmetries in charmless $B$ decays. We use 34.6 fb$^{-1}$ from electron-positron collisions collected in 2019 and 2020 at the $\Upsilon(4S)$ resonance.

The results are compatible with known values and show performance already comparable with Belle.
\end{abstract}

\section{Introduction}
Charmless $B$ decays are transitions where a $b$ quark decays into quarks other than $c$. They are central to the scientific program of Belle~II as they provide unique access to parameters of the weak interactions that are particularly sensitive to non-Standard-Model physics, including the quark-mixing phase $\alpha/\phi_2 = {\rm arg}\left[{V_{tb}^* V_{td}}/{(-V_{ub}^* V_{ud})}\right]$ \cite{kou}, the $K\pi$ isospin sum-rule \cite{kou, gronau}, and charge-parity (CP) violating asymmetries localized in the phase space of three-body $B$ decays \cite{kou}. 

Belle II~\cite{kou} is a detector that reconstructs the products of energy-asymmetric 7-on-4 GeV  electron-positron collisions produced by SuperKEKB, a collider located at the KEK laboratory in Japan. SuperKEKB produces boosted $\Upsilon(4S)$ mesons to achieve $B$ meson decay lengths resolvable with current vertex detectors. A nano-beam collision scheme allows for luminosities sufficiently high to accumulate $50$~ab$^{-1}$ by 2030, about 40 times the amount collected by $B$ factories to date.
Belle II comprises several subdetectors arranged in a cylindrical geometry around the interaction region. Most relevant for this work are the silicon vertex detector, which samples the trajectories of charged particles in the vicinity of the interaction region (impact-parameter resolution of 20~$\mu$m); a drift chamber, which measures charged-particle momenta (transverse momentum resolution $\sigma(p_T)/p_T^2 \approx 0.5\%/[{\rm GeV}/c]$), charge, and specific ionization, two Cherenkov particle-identification detectors (5$\sigma$ kaon/pion separation), and an electromagnetic calorimeter, which measures the energy $E$ of electrons and photons ($\sigma_E/E=1.6\%\--8\%$).
The sample used in this work corresponds to an integrated luminosity of 34.6 fb$^{-1}$~\cite{lumi} and was collected at the $\Upsilon(4S)$ resonance as of May 14, 2020.

We perform first measurements in charmless two-body $B$ decays $B^0\to K^+\pi^-$, $B^+\to K^+\pi^0$, $B^+\to K^{0}_{\rm S}\pi^+$, $B^0\to K^{0}_{\rm S}\pi^0$, $B^0\to \pi^+\pi^-$, $B^+\to \pi^+\pi^0$ and three-body decays $B^+\to K^+K^-K^+$, $B^+\to K^+\pi^-\pi^+$~\cite{charmless1}. These benchmark multiple and diverse experimental capabilities, including vertexing, charged- and neutral-particle reconstruction, particle identification, and background suppression. Charge-conjugate processes are implied, unless specified otherwise.

We reconstruct the decays, optimize their selections, and extract the signal yields by fitting the difference $\Delta E$ between expected and observed candidate energy. We then correct yields for absolute (charge-specific) efficiencies from simulation (control data) to obtain the branching fractions (CP-violating asymmetries).

\section{Analysis}
The principal challenge is to isolate a signal lacking final-state leptons and intermediate resonances from a $10^{5}$ times larger background dominated by random combinations of particles produced in $continuum$ $e^+e^-\to q\bar{q}$ ($q=u,d,s,c$) events.

We use two strongly discriminating variables, the beam-energy-constrained mass, $M_{\rm bc} \equiv \sqrt{s/(4c^4) - (p^{*}_B/c)^2}$, and energy difference, $\Delta E \equiv E^{*}_{B} - \sqrt{s}/2$, where $\sqrt{s}$ is the collision energy, $E^{*}_{B}$ and $p^{*}_{B}$ are the reconstructed energy and momentum of $B$ meson candidates, all in the $\Upsilon(4S)$ frame. While $M_{\rm bc}$ discriminates between signal, which peaks at the $B$ meson mass, and $continuum$, which has a nearly uniform distribution, $\Delta E$ discriminates also correctly reconstructed $B$ mesons, which peak at $\Delta E = 0$, from the mis-reconstructed ones, whose peak are shifted. For further discrimination, we use a binary boosted decision-tree classifier that combines non-linearly about 30~variables associated with the topology and momentum-energy flow of the event,  kinematics, and the flavor content of the partner $B$ meson. We train the classifier to identify discriminating features using unbiased simulated samples. Only properly simulated inputs, as verified in control samples, are used.

To reconstruct $B$ candidates, we first form final-state particle candidates by applying a baseline selection. Then we combine these candidates in kinematic fits consistent with the target topologies. 

We optimize each selection by varying simultaneously the requirements on the classifier output and charged-particle identification to maximize S/$\sqrt{\textrm{S+B}}$, where S and B are signal and background yields in simulated samples. The $\pi^0$ selection is optimized using $B^+\to\overline{D}^0(\to K^+\pi^-\pi^0)\pi^+$ decays reconstructed in data, due to known mismodelings in simulation. We restrict samples to one candidate per event, chosen randomly. 
Simulation shows 0-6$\%$ $\textit{self-cross-feed}$, i.e., misreconstructed signal candidates formed by misidentified signal particles or combinations of signal and non-signal particles. We include the self-cross-feed component in our fit model by fixing its proportions to the expectations from simulation.

We use simulation to suppress contamination from peaking backgrounds in three-body decays. We exclude the two-body masses corresponding to $D^0$, $\eta_c$, and $\chi_{c1}$ decays for $B^+\to K^+K^-K^+$ and $D^0$, $\eta_c$, $\chi_{c1}$, $J/\psi$, and $\psi(2S)$ decays for $B^+\to K^+\pi^-\pi^+$ decays.

\begin{figure}[!htb]
\begin{center}
\includegraphics[width=5.65cm]{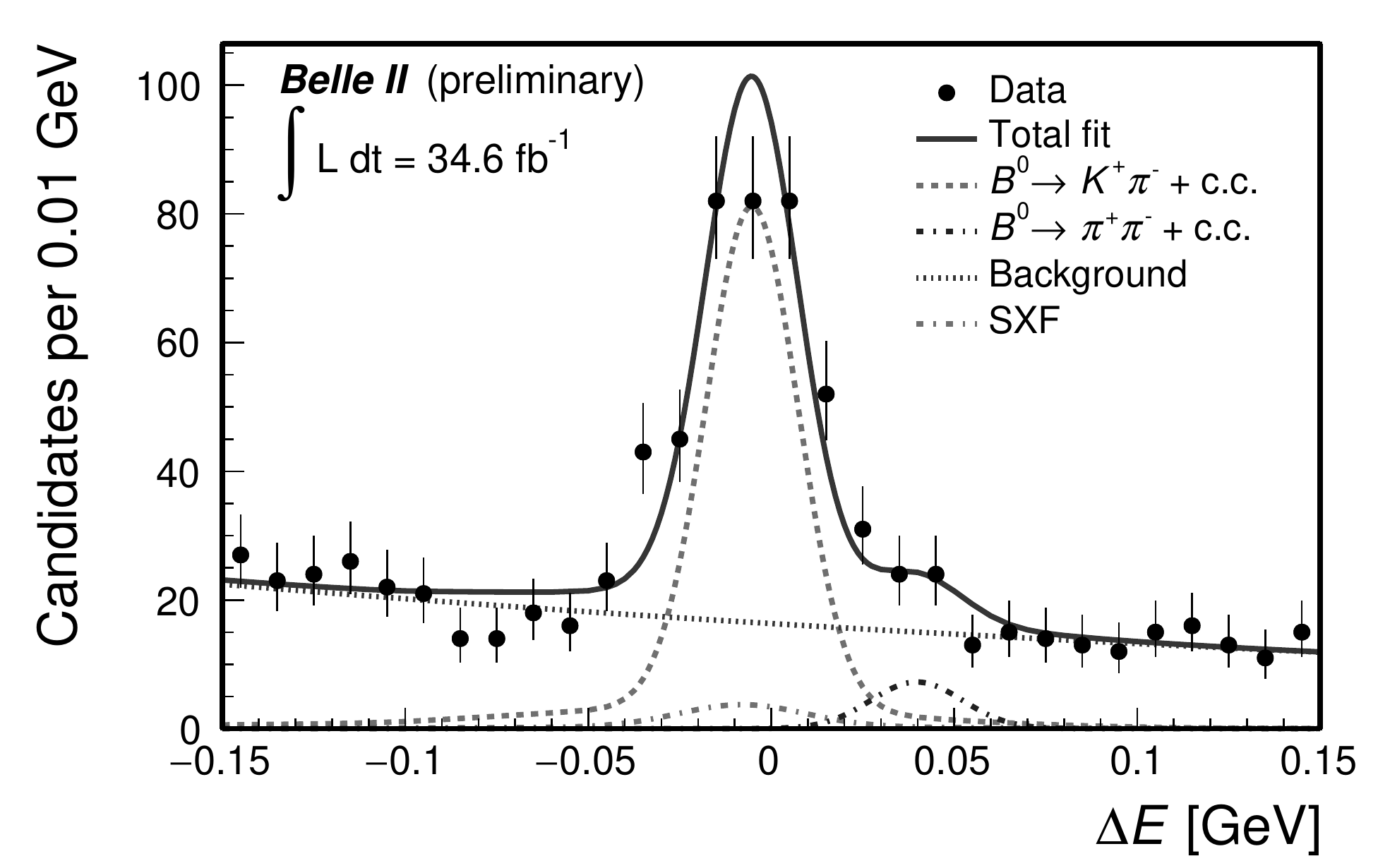}
\includegraphics[width=5.65cm]{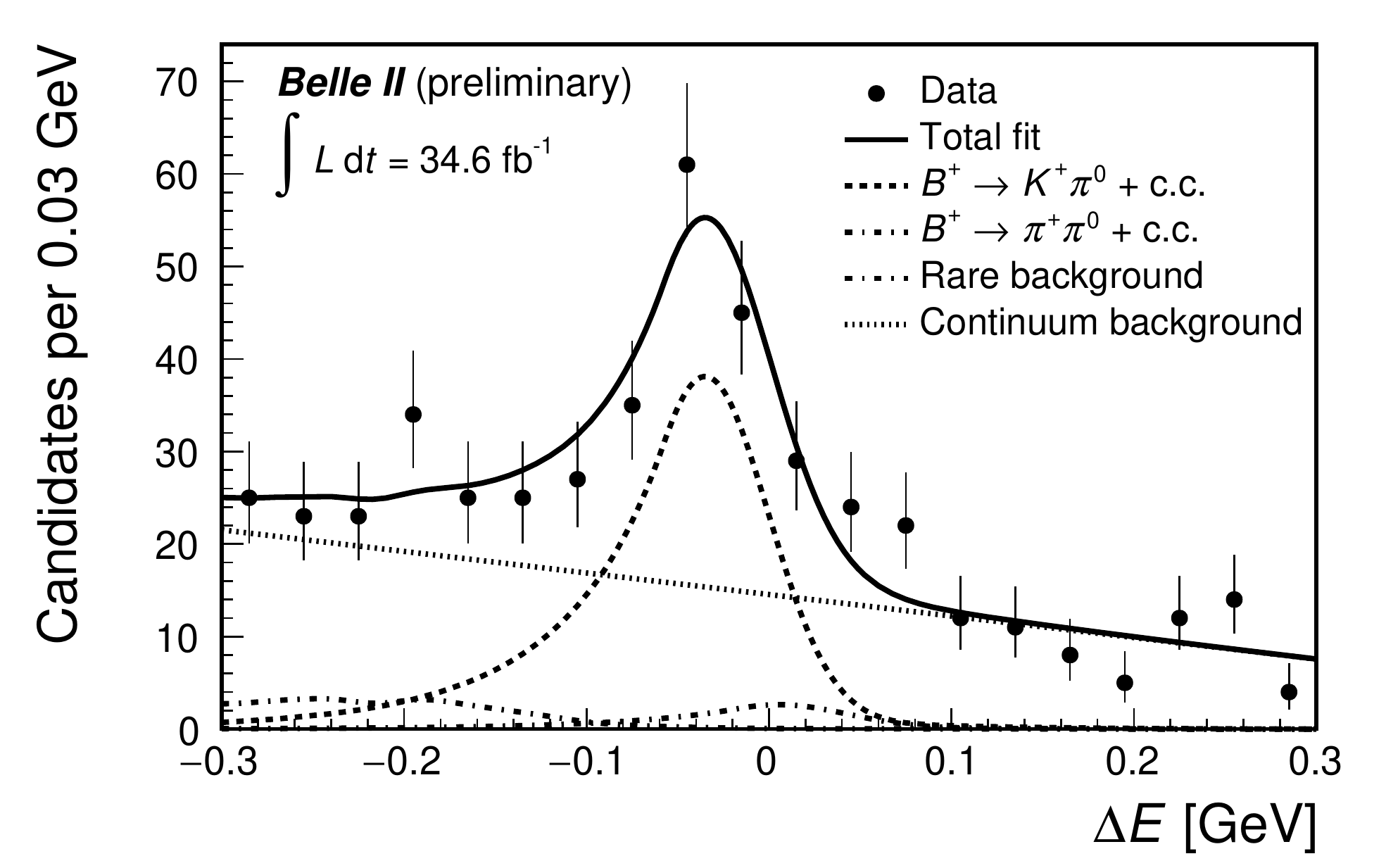}
\includegraphics[width=5.65cm]{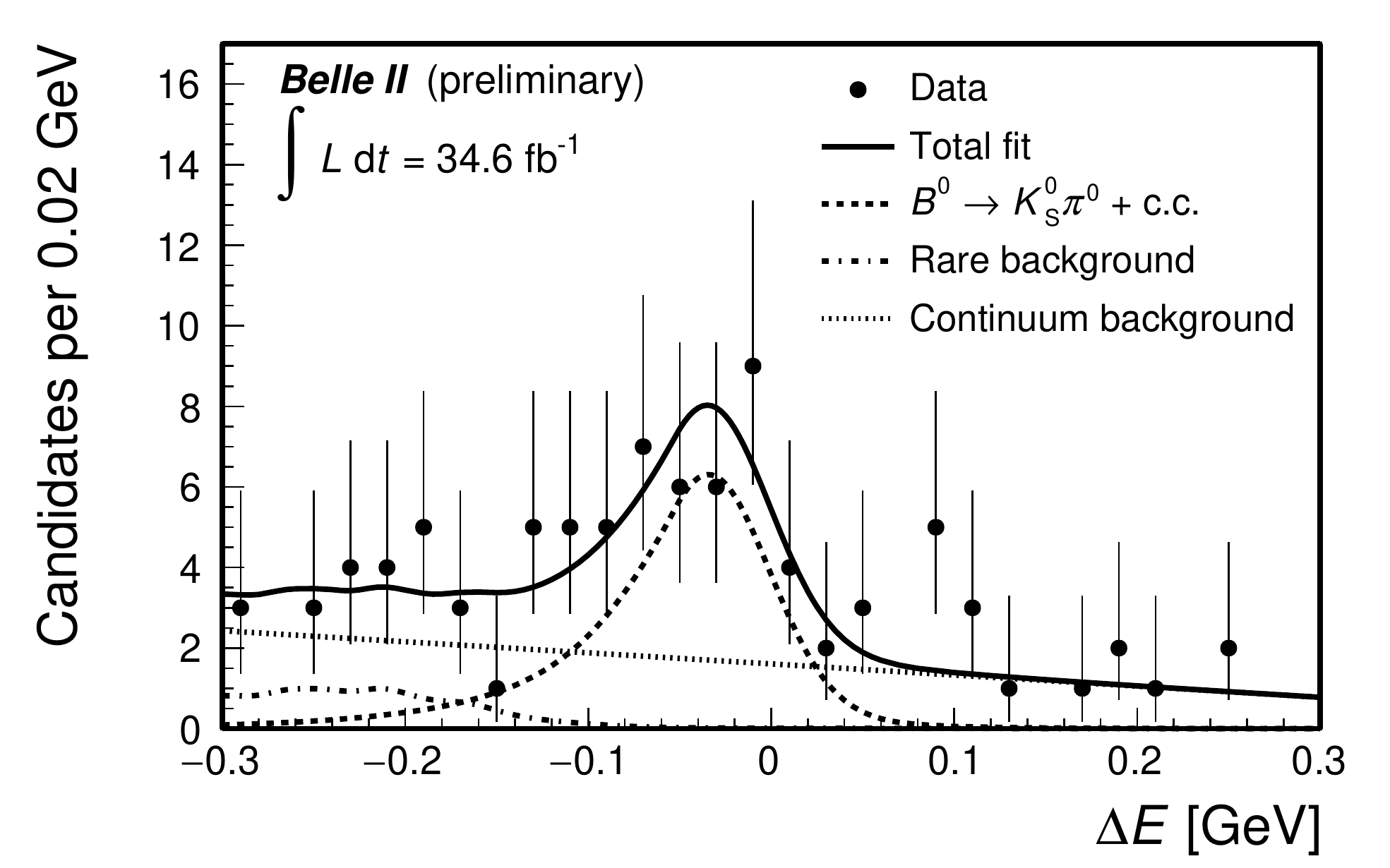}
\includegraphics[width=5.65cm]{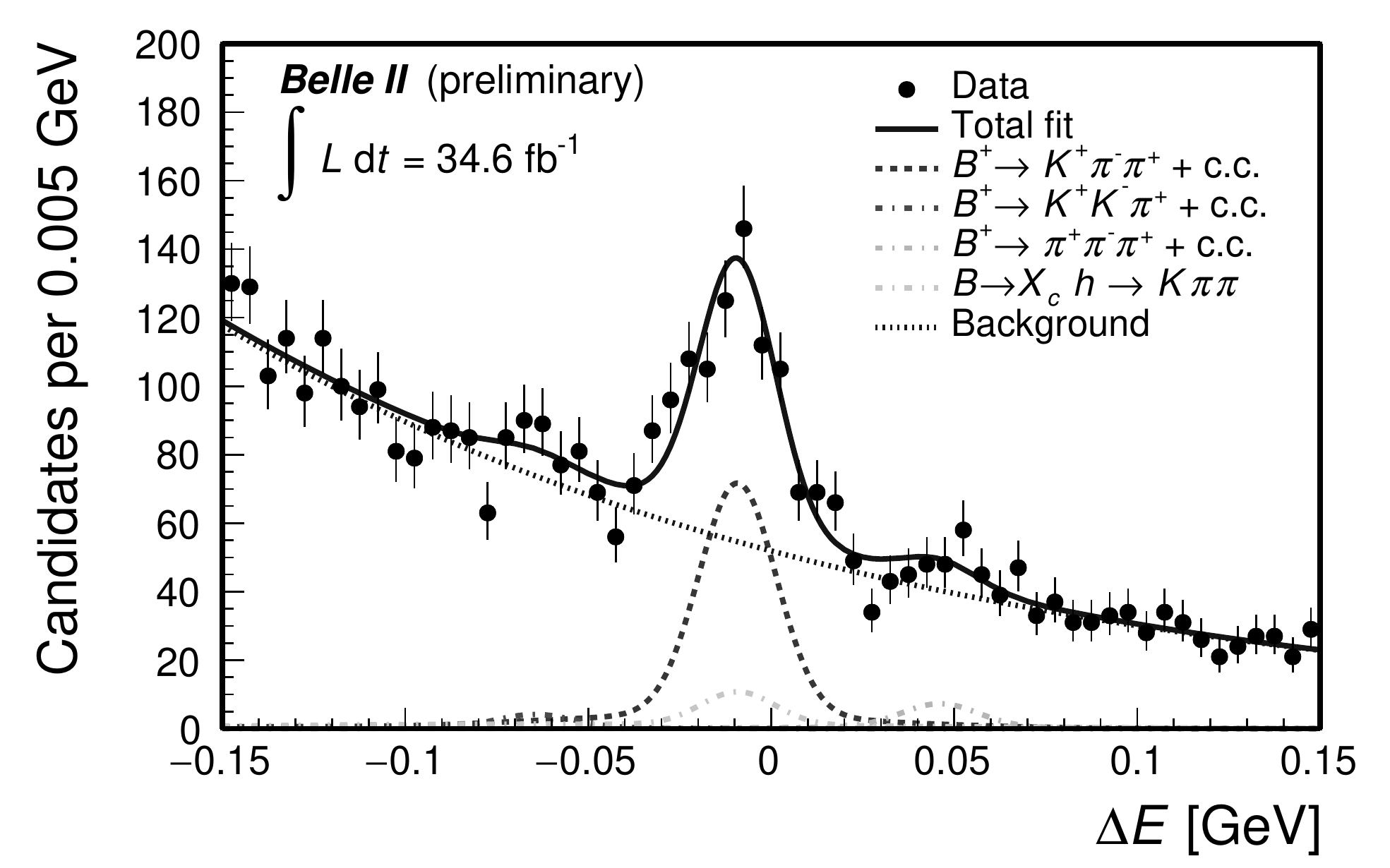}
\end{center}
 \caption{Distributions of $\Delta E$ for (top left) $B^{0}\to K^{+}\pi^{-}$, (top right) $B^{+}\to K^{+}\pi^{0}$, (bottom left) $B^{0}\to K^{0}_{\rm S}\pi^{0}$, and (bottom right) $B^{+}\to K^{+}\pi^{-}\pi^{+}$ candidates reconstructed in 2019--2020 Belle II data with fit projections overlaid. The "SXF" label indicates self-cross-feed.}
 \label{fig:Kpi}
\end{figure}

We determine signal yields from maximum likelihood fits of the unbinned $\Delta E$ distributions of candidates restricted to $M_\textrm{bc}$ $>$ 5.27 GeV/$c^2$ and $|\Delta E|$ $<$ 0.15 GeV (Fig.~\ref{fig:Kpi}). Fit models are obtained empirically from simulation, with the only additional flexibility being a possible shift of the signal-peak position, which is determined in data.

We determine each branching fraction as $\mathcal{B} = N/(2\varepsilon N_{B\overline{B}})$, where $N$ is the signal yield obtained from the fit; $\varepsilon$ is the selection efficiency, determined from simulation and validated in control data; $N_{B\overline{B}}$ is the number of produced $B\bar{B}$ pairs, determined from the measured integrated luminosity, the $e^+e^-\to\Upsilon(4S)$ cross section, and the $\Upsilon(4S)\to B^0\bar{B}^0$ branching fraction \cite{bphys}.

\begin{figure}[htb]
 \centering
 \includegraphics[width=5.65cm]{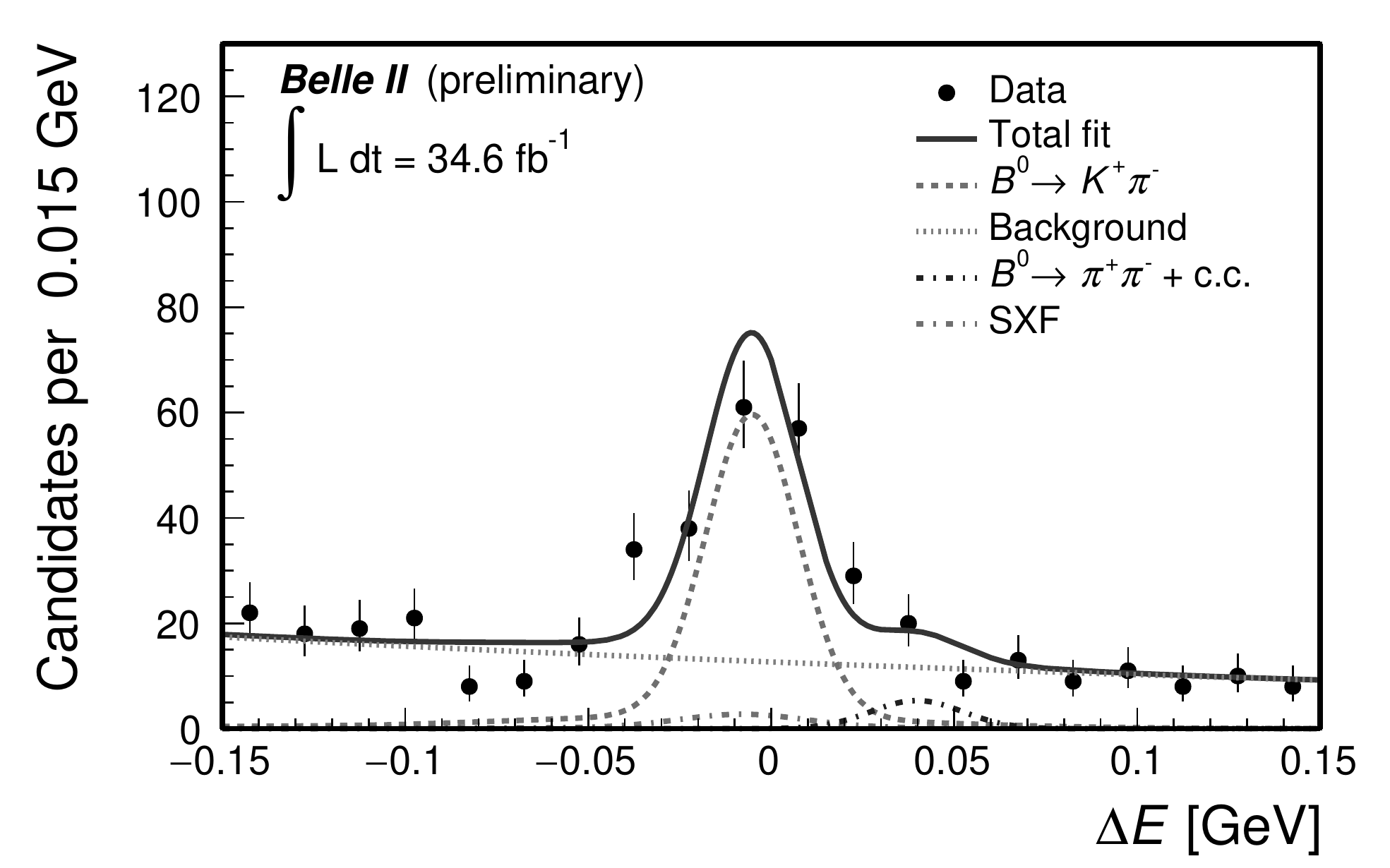}
 \includegraphics[width=5.65cm]{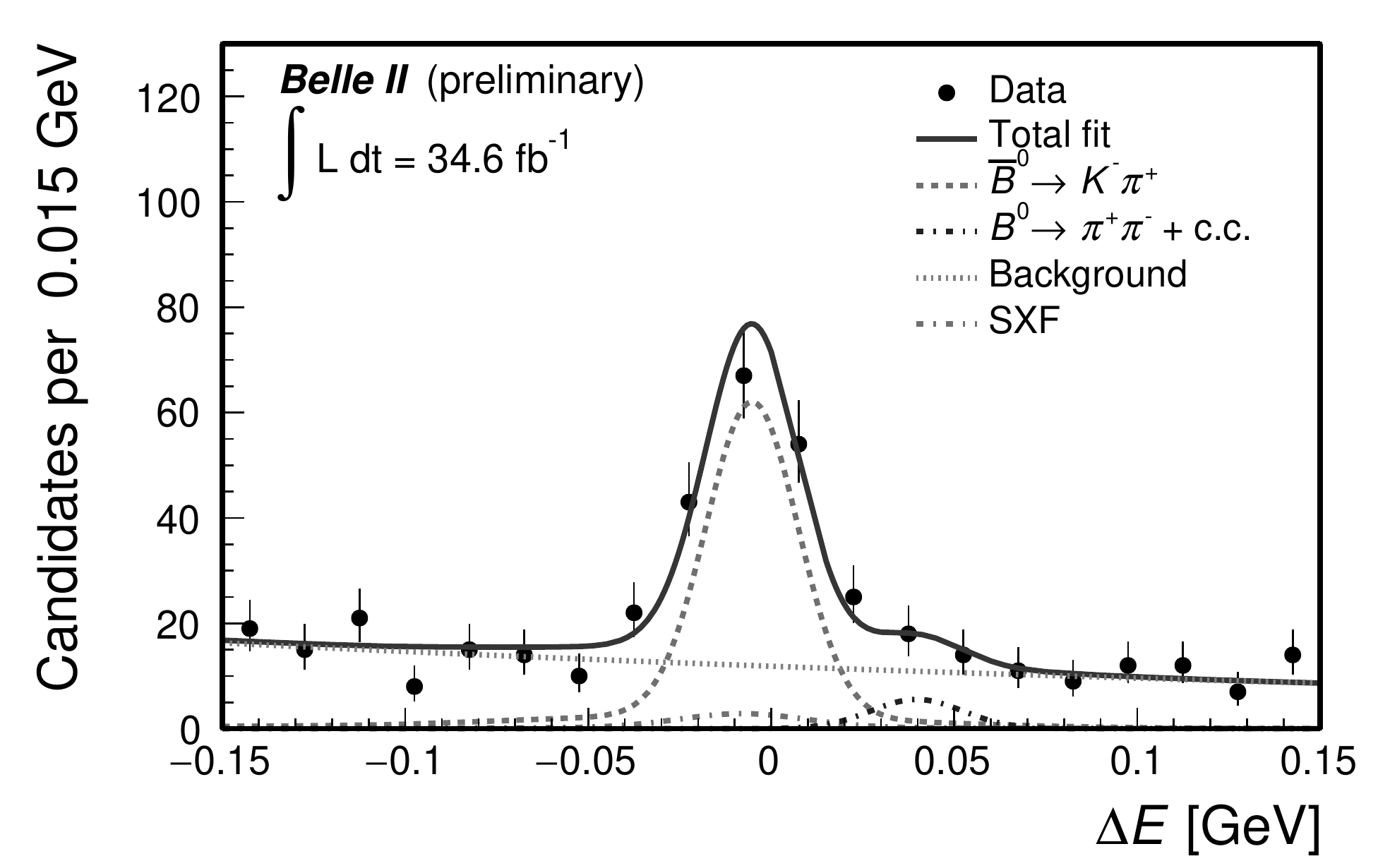}
 \caption{Distributions of $\Delta E$ for (left) $B^0 \to K^+\pi^-$ and (right) $\overline{B}^0 \to K^-\pi^+$ candidates reconstructed in 2019--2020 Belle~II data with fit projections overlaid.}
 \label{fig:ACP_K+pi-}
\end{figure}

We also measure CP-violating asymmetries $\mathcal{A}_{\rm CP} = \mathcal{A} - \mathcal{A}_{\rm det}$, where $\mathcal{A}$ is the observed charge-specific signal-yield asymmetry, and $\mathcal{A}_{\rm det}$ is the instrumental asymmetry due to differences in interaction or reconstruction probabilities between opposite-charge hadrons~\cite{Raiz:2020tcz}. We determine $\mathcal{A}$ from a simultaneous non-extended likelihood fit of the unbinned $\Delta E$ distributions of bottom and antibottom candidates decaying in flavor-specific final states (Fig.~\ref{fig:ACP_K+pi-}). We evaluate the instrumental asymmetries $\mathcal{A}_{\rm det}(K^+\pi^-)$, $\mathcal{A}_{\rm det}(K^0_{\rm S}\pi^+)$, and $\mathcal{A}_{\rm det}(K^+)$, by measuring the charge-asymmetry in abundant samples of $D^0\to K^-\pi^+$ and $D^+\to K^0_{\rm S}\pi^+$ decays, as well as using Ref.~\cite{Raiz:2020tcz}, obtaining asymmetries of 1--2$\%$ with 2$\%$ uncertainties.
Finally, we assess systematic effects associated with shape modelling, $\pi^0$ reconstruction, tracking, and particle identification. 

Table~\ref{tab:res1} lists the results. All are compatible with known values, and the yields per inverse femtobarn and purities are on par with the best Belle performances.

\begin{table}[h]
\centering
\footnotesize
\caption{Results. The first contribution to the uncertainty is statistical, the second is systematic. A 0.5 factor accounts for the $K^0\to K^0_{\rm S}$ probability.}
\begin{tabular}{l c c c}
\hline\hline
 Decay mode & Branching fraction $\times~10^{-6}$ & CP-violating asymmetry \\
 \hline
 $B^{0}\to K^{+}\pi^{-}$ &  $18.9 \pm 1.4 \pm 1.0$ & $0.030 \pm 0.064 \pm 0.008$ \\[0.06cm]
 $B^+ \to K^+\pi^0$ & $12.7 ^{+2.2}_{-2.1}\pm 1.1$ & $0.052 ^{+0.121}_{-0.119}\pm 0.022$ \\[0.06cm]
 $B^+ \to K^0\pi^+$ & $21.8 ^{+3.3}_{-3.0} \pm 2.9$ & $-0.072 ^{+0.109}_{-0.114} \pm 0.024$ \\[0.06cm]
 $B^0 \to K^0\pi^0$ & $10.9^{+2.9}_{-2.6} \pm 1.6$ & -  \\[0.06cm]
 $B^0 \to \pi^+\pi^-$ & $5.6 ^{+1.0}_{-0.9} \pm 0.3$ & -  \\[0.06cm]
 $B^+ \to \pi^+\pi^0$ & $5.7 \pm 2.3\pm 0.5$ & $-0.268 ^{+0.249}_{-0.322}\pm 0.123$ \\[0.06cm]
 $B^+ \to K^+K^-K^+$  & $32.0 \pm 2.2 \pm 1.4$ & $-0.049 \pm 0.063 \pm 0.022 $ \\[0.06cm]
 $B^+ \to K^+\pi^-\pi^+$ & $48.0 \pm 3.8\pm 3.3$ & $-0.063 \pm 0.081 \pm 0.023$ \\[0.06cm]
\hline\hline
\end{tabular} 

\label{tab:res1}
\end{table}

\section{Summary}

We report on first measurements of branching fractions and CP-violating charge asymmetries in charmless $B$ decays at Belle II. We use a sample of 2019 and 2020 data corresponding to 34.6 fb$^{-1}$ of integrated luminosity. We determine branching fractions for the $B^0\to K^+\pi^-$, $B^+\to K^+\pi^0$, $B^+\to K^{0}_{\rm S}\pi^+$, $B^0\to K^{0}_{\rm S}\pi^0$, $B^0\to \pi^+\pi^-$, $B^+\to \pi^+\pi^0$, $B^+\to K^+K^-K^+$, and $B^+\to K^+\pi^-\pi^+$ decays, and direct CP-violating asymmetries for the decays $B^0\to K^+\pi^-$, $B^+\to K^+\pi^0$, $B^+\to K^{0}_{\rm S}\pi^+$, $B^+\to \pi^+\pi^0$, $B^+\to K^+K^-K^+$, and $B^+\to K^+\pi^-\pi^+$. The results agree with known values and show detector performance already comparable with the best Belle results, indicating a good understanding of the detector and offering a reliable basis to assess projections for future reach.

\end{document}